\documentclass[%
 reprint,
nofootinbib,
 amsmath,amssymb,
 aps,
floatfix,
prl,
]{revtex4-2}

\usepackage[ruled,vlined,linesnumbered,noalgohanging]{algorithm2e}

\SetAlCapSty{myalcaptitle}

\SetAlCapNameSty{myalcap}
\SetAlCapHSkip{0pt}
\SetAlgoCaptionSeparator{.}
\SetAlgoCaptionLayout{raggedright}
\SetAlgoHangIndent{0pt}

\usepackage{graphicx}
\usepackage{bm}
\usepackage{hyperref}
\usepackage{xspace}
\usepackage[capitalise]{cleveref}
\usepackage{lmodern}
\usepackage{microtype}
\usepackage{fontawesome}
\usepackage[T1]{fontenc}
\usepackage{siunitx}

\hypersetup{
    colorlinks=true,
    allcolors=[rgb]{0.26,0.41,0.88},
}

\newcommand{\nlive}{n_\text{live}}
\newcommand{\niter}{n_\text{iter}}

\newcommand{\pvalue}{\text{\textit{p-}value}\xspace}
\newcommand{\pvalues}{\text{\pvalue{}s}\xspace}
\newcommand{\Pvalue}{\text{\textit{P-}value}\xspace}
\newcommand{\Pvalues}{\text{\Pvalue{}s}\xspace}

\newcommand{\like}{\mathcal{L}}

\newcommand{\pg}[2]{p\mathopen{}\left(#1\,\rvert\, #2\right)\mathclose{}}
\newcommand{\Pg}[2]{P\mathopen{}\left(#1\,\rvert\, #2\right)\mathclose{}}

\newcommand{\intd}{\text{d}}
\newcommand{\sampleParams}{\mathbf{x}}
\newcommand{\modelParams}{\boldsymbol{\theta}}
\newcommand{\param}{x}

\newcommand{\MN}{\textsc{MultiNest}\xspace}
\newcommand{\PC}{\textsc{PolyChord}\xspace}

\newcommand{\ee}{\mathrm{e}}

\usepackage[usenames]{xcolor}

\newcommand{\refcite}{\cite}

\newcommand{\wordlimit}[1]{} 

\begin{document}

\title{Nested sampling for frequentist computation: fast estimation of small \pvalues}

\author{Andrew Fowlie}
\affiliation{Department of Physics and Institute of Theoretical Physics, Nanjing Normal University, Nanjing, Jiangsu 210023, China}
\email{andrew.j.fowlie@njnu.edu.cn}
\author{Sebastian Hoof}
\affiliation{Institut f\"{u}r Astrophysik, Georg-August-Universit\"{a}t {G\"{o}ttingen},\\Friedrich-Hund-Platz~1, 37077~{G\"{o}ttingen}, Germany}
\author{Will Handley}
\affiliation{Cavendish Laboratory \& Kavli Institute for Cosmology, University of Cambridge,\\ JJ Thomson Avenue, Cambridge, CB3~0HE, United Kingdom}

\begin{abstract}
    We propose a novel method for computing \pvalues based on nested sampling (NS) applied to the sampling space rather than the parameter space of the problem, in contrast to its usage in Bayesian computation. The computational cost of NS scales as $\log^2{1/p}$, which compares favorably to the $1/p$~scaling for Monte Carlo (MC) simulations. For significances greater than about $4\sigma$ in both a toy problem and a simplified resonance search, we show that NS requires orders of magnitude fewer simulations than ordinary MC estimates. This is particularly relevant for high-energy physics, which adopts a $5\sigma$ gold standard for discovery. We conclude with remarks on new connections between Bayesian and frequentist computation and possibilities for tuning NS implementations for still better performance in this setting. \href{https://github.com/andrewfowlie/ns_for_p_values}{\faGithub}
\end{abstract}

\maketitle

\section{Introduction}\label{sec:intro}

For decades, \pvalues have played a vital role in the discovery of new phenomena in high-energy physics~(HEP)~\citep{Lyons:1986em,Cowan:2013pha,Cranmer:2015nia,Cousins:2018tiz,Zyla:2020zbs,junk} as well as many other fields and disciplines. A~\pvalue is the probability under a null hypothesis of observing results that are at least as ``extreme'' as the observed data. When the observed \pvalue is smaller than a pre-specified threshold $\alpha$ we reject the null hypothesis and claim a discovery.

As \wordlimit{brought to the attention of the public} popularized by the Higgs boson discovery~\citep{Aad:2012tfa,Chatrchyan:2012ufa}, HEP usually requires $\alpha \sim 10^{-7}$, also known as the ``$5\sigma$ rule''~\citep{Lyons:2013yja}.\footnote{In HEP, it is conventional to convert \pvalues into significances by the rule $Z = \Phi^{-1}(1 - p)$, where $\Phi$ is the standard normal cumulative distribution function.}
Establishing a discovery thus requires the computation of $p \lesssim 10^{-7}$. This exposes difficulties in standard approaches to computing \pvalues, including the look-elsewhere effect~\cite[see e.g.,][]{Algeri:2016gtj}, broken assumptions in popular asymptotic results~\citep[see e.g.,][]{Cowan:2010js}
and the computational cost of computing \pvalues through Monte Carlo (MC) simulations. 
To overcome these problems, semi-analytic asymptotic formulae were developed alongside the Higgs searches~\cite[e.g., Gross--Vitells][]{Gross:2010qma}.

However, as discussed in the reviews \refcite{Cranmer:2015nia,Cousins:2018tiz,Zyla:2020zbs}, MC simulations are often unavoidable, as the asymptotic formulae make assumptions that often do not hold or that are difficult to check\wordlimit{, as discussed in detail in}~\refcite{2020NatRP...2..245A}. Straightforward examples include small sample sizes that cannot justify taking the asymptotic limit or when the hypotheses under consideration are not nested.
\wordlimit{Since it is impossible to select and summarize prominent examples from the diverse range of disciplines using \pvalues, we only mention some recent examples in physics.} For example, MC simulations were used by ATLAS and CMS at the LHC in e.g., searches for deviations from the Standard Model (SM) of particle physics~\cite{1807.07447,2010.02984}, searches for supersymmetry~\cite{1405.3961} and measurements of the Higgs boson's properties~\cite{1212.6639,1405.3961}. Outside collider physics, they were used in searches for dark matter by the XENON collaboration~\cite{2006.09721,2105.00599} and in astronomy by Fermi-LAT~\cite{1611.03184} and IceCube~\cite{2107.08159}. Lastly, MC simulations are used in global fits in particle physics~e.g., \refcite{1508.05951,2007.05517,2104.05631,1311.1822,1506.07685,2008.06083}, including \textsc{Gfitter} fits of the SM~\cite{1407.3792}.

In this \textit{Letter} we present a novel technique for computing global or local \pvalues based on nested sampling (NS) that in some regimes performs exponentially better than MC simulations.
\wordlimit{In \cref{sec:pvalues} we describe our technique and compare it with MC simulations. We then assess the usability of our algorithm for benchmark cases in \cref{sec:examples}. Finally, we conclude in  \cref{sec:conc}, including a brief discussion of limitations of our approach and an explanation of its success by rephrasing the problem in terms of \emph{compression}.} A basic implementation of our algorithm, including the examples provided in this \textit{Letter}, is available on Github~\cite{code}.

\section{\pvalues}\label{sec:pvalues}
Mathematically, \pvalues can be defined as the probability that a test-statistic (TS) $\lambda$ is at least as great as the \emph{observed} TS $\lambda^\star$ assuming that the null hypothesis $H_0$ is true,
\begin{equation}\label{eq:pval_def}
     p =\Pg{\lambda \ge \lambda^\star}{H_0} \, .
\end{equation}
The task at hand is therefore to determine a tail probability from the sampling distribution of~$\lambda$ under~$H_0$. \wordlimit{Typically this involves generating simulated sets of data $\sampleParams$ from the \emph{sampling space} of the null hypothesis, and using these to estimate the fraction for which $\lambda(\sampleParams)\ge\lambda^\star$ holds true.}

\subsection{\Pvalues from Monte Carlo simulations}\label{sec:mc}
Using MC simulations to \wordlimit{correctly} determine the sampling distribution requires no asymptotic assumptions and \wordlimit{generally} provides a robust and reliable \pvalue estimate~\refcite{Algeri:2016gtj}. These estimates obey binomial statistics since for each simulation either $\lambda \geq \lambda^\star$ or not.
The \pvalue can then be estimated by $\hat{p} = m/n$ for $m$ occurrences of $\lambda \geq \lambda^\star$ from $n$ simulations. Evidently, a meaningful estimate requires at least $1/p$ simulations as the fractional error on~$\hat{p}$ is of order of the Wald estimate,
\begin{equation}\label{eq:mc_pval_err}
    \frac{\Delta p}{p} = \sqrt{\frac{1 / p}{n}} \, .
\end{equation}
See \refcite{10.1214/ss/1009213286} for further discussion of errors for binomial parameters.

\subsection{\Pvalues from nested sampling}\label{sec:ns}

\begin{figure*}
    \centering
    \includegraphics[width=7.0in]{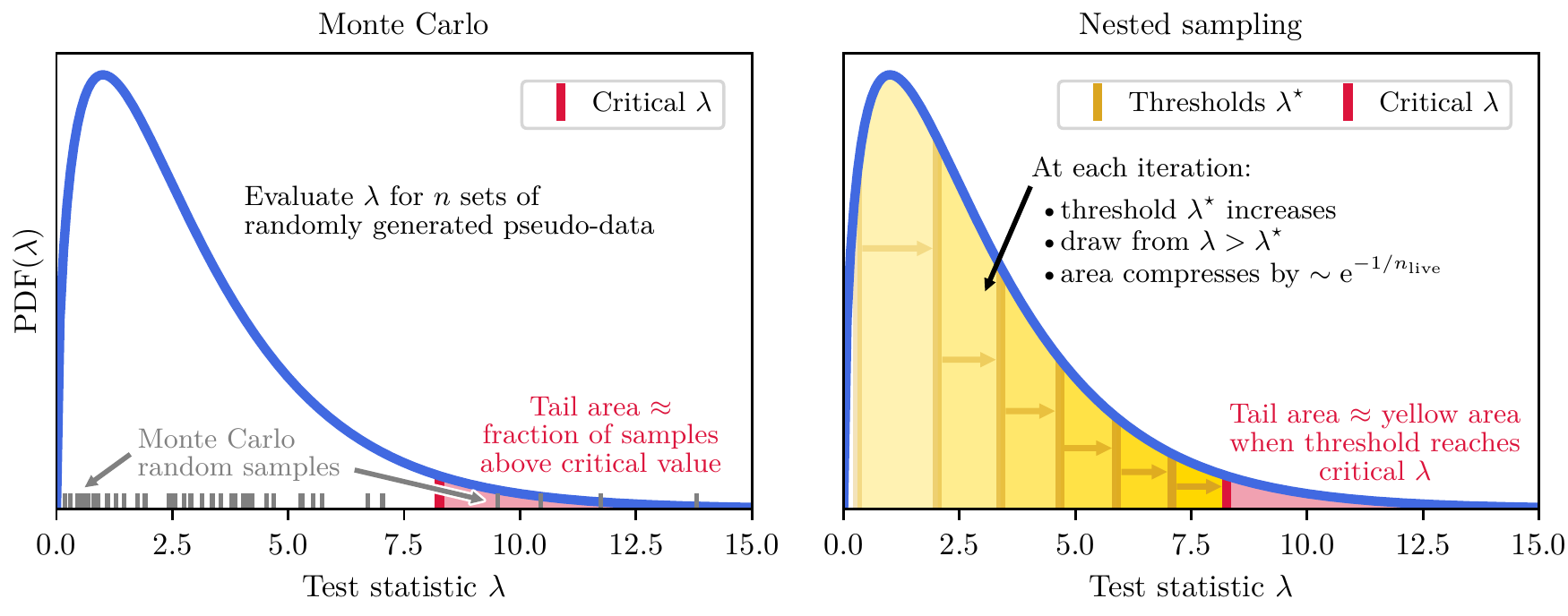}
    \caption{Schematic illustration of MC~(\textit{left}) and NS~(\textit{right}) methods for computing \pvalues. The blue curve represents the probability distribution of the test statistic~$\lambda$. The observed, critical value of~$\lambda$ is indicated by a thick, red, vertical line and the area beyond it~(red shaded region) is the \pvalue. The shaded yellow region in the right panel illustrates the compression of the area under the curve at each step of the NS algorithm~(progressively darker shading).}
    \label{fig:ill}
\end{figure*}

The nested sampling~(NS) algorithm~\citep{2004AIPC..735..395S,Skilling:2006gxv} has primarily enjoyed success as a tool for computing the Bayesian evidence~(see e.g.~\refcite{Kass:1995loi}), which is estimated through an identity involving the volume variable
\begin{equation}\label{eq:X}
    X(\like^\star) = \int_{\like(\modelParams) > \like^\star} \pi(\modelParams) \, \intd\modelParams \, ,
\end{equation}
where $\pi$ is the prior distribution, $\like$ is the likelihood and $\modelParams$ are the model parameters.
The threshold $\like^\star$ increases at each iteration of the algorithm. 
A scheme to estimate $X$ 
lies at the heart of NS. See \refcite{2021arXiv210109675B} for a recent review.

To understand why this is useful for \pvalue calculations, let us re-interpret \cref{eq:X} in a frequentist context. Rather than thinking of it as an integral over the parameter space, consider it \wordlimit{as} an integral over the sampling space. To give a concrete example, if we wish to simulate five Gaussian measurements, the sampling space would be five-dimensional and $\modelParams$ would consist of the five simulated Gaussian measurements $\sampleParams$ rather than model parameters. We thus re-write \cref{eq:X} as 
\begin{equation}\label{eq:p_value_X}
   p = \Pg{\lambda > \lambda^\star}{H_0} = \int_{\lambda(\sampleParams) > \lambda^\star} \pg{\sampleParams}{H_0} \, \intd\sampleParams \, ,
\end{equation}
which allows us to use NS to estimate \pvalues. A comparison of \cref{eq:X,eq:p_value_X} reveals that the pseudo-data~$\sampleParams$ play the role of the model parameters~$\modelParams$, the sampling distribution of the pseudo-data $\pg{\sampleParams}{H_0}$ plays the role of the prior and, lastly, the TS $\lambda(\sampleParams)$ plays the role of the likelihood. Thus in the Bayesian setting, the volume variable in \cref{eq:X} is the fraction of the prior in which the likelihood exceeds a threshold, whereas in the frequentist one in \cref{eq:p_value_X}, it is the fraction of the sampling distribution in which the TS exceeds a threshold.

The NS scheme for a statistical estimate of the volume variable works as follows. First, we draw $\nlive \sim \mathcal{O}(\numrange{100}{1000})$ ``live points'' from the sampling space. The parameter $\nlive$ acts as a resolution parameter, such that the runtime scales approximately as $\nlive$, the uncertainty as $1 / \sqrt{\nlive}$ and the chances of missing modes in multi-modal problems as $1/\nlive$. At each subsequent iteration, the live point with the least extreme TS is replaced by one with a more extreme TS drawn from the sampling space (i.e.\ a draw from the constrained prior). This continues until iteration $\niter$, when the least extreme TS among the live points exceeds the observed TS. \wordlimit{We denote the iteration at which that occurs by $\niter$.} We estimate the \pvalue using the usual NS estimate of the volume,\footnote{This follows from $\ee^{\langle\log X\rangle}$, i.e., from an unbiased estimator of $\log X$. 
} 
\begin{equation}\label{eq:p_NS}
   p = X(\lambda^\star) \simeq \prod_{i=1}^{\niter} \ee^{-1/\nlive} = \ee^{-\niter / \nlive} \, .
\end{equation}
Thus NS decomposes a tiny \pvalue into a product of $\niter$ moderate factors. This algorithm is shown schematically in \cref{algo:ns_pvalues}.

\begin{algorithm}
\caption{Schematic nested sampling algorithm for estimating \pvalues.\label{algo:ns_pvalues}}
 \SetAlgoLined
 Draw $\nlive$ sets of pseudo-data from the sampling distribution --- the live points\;
 Initialize $\niter = 0$\;
 \Repeat{
    $\lambda_\mathrm{min} \ge \lambda^\star$
    }{
    $\niter = \niter + 1$\;
    Find the minimum TS $\lambda_\mathrm{min}$ amongst the live points\;
    Replace live point corresponding to $\lambda_\mathrm{min}$ by one drawn from the sampling distribution subject to $\lambda > \lambda_\mathrm{min}$\;
    }
\KwRet{Estimate of $p = \ee^{-\niter / \nlive}$}\;
\end{algorithm}

\wordlimit{In more practical terms, we initialize NS with the TS in place of the likelihood and the sampling distribution in place of the prior. As the NS algorithm runs, the threshold in \cref{eq:p_value_X} increases monotonically. We stop once we reach a TS of~$\lambda^\star$, where we may choose~$\lambda^\star$ as the observed TS in an experiment or some arbitrary value for the purpose of calibrating the distribution of the TS before performing an experiment.} 

In the NS algorithm, $\niter$ is approximately Poisson-distributed with expectation value and variance both equal to $\nlive \log 1 / p$. From \cref{eq:p_NS} we see that $\log p$ follows a normal distribution with error
\begin{equation}\label{eq:ns_pval_err}
 \frac{\Delta p}{p} \approx \Delta \log p = \frac{\Delta \niter}{\nlive} = \sqrt{\frac{\log 1/p}{\nlive}} \, ,
\end{equation}
with the approximation holding when $\Delta p / p \lesssim 1$. However, the number of TS evaluations required in an NS run $n_\text{eval}$ must be proportional to the number of iterations before stopping. In fact, from the known properties of NS, 
\begin{equation}
n_\text{eval} = \niter / \epsilon = (\nlive  \log 1 / p) / \epsilon \, ,   
\end{equation}
where $\epsilon$ denotes a problem-specific efficiency factor.\footnote{This neglects the (generally negligible) $\nlive$ evaluations required for the initial live points.} Thus we actually have
\begin{equation}
\frac{\Delta p}{p} \approx \Delta \log p =  \sqrt{\frac{\log^21/p}{\epsilon \, n_\text{eval}}} \, .
\end{equation}
\wordlimit{The theoretical foundations for NS in this setting are further strengthened by the connection to a form of subset simulation~\cite{Au_2001,beck2015rare}, a technique from rare event sampling that achieves similar $\log^r{1/p}$ scaling with $r = \numrange{2}{3}$.}

To compare the theoretical performances of MC and NS, we consider the ratio of the number of evaluations required to achieve a fixed relative error on the \pvalue. Note that the TS could be based on the profiled likelihood such that each evaluation could involve a complicated minimization over the actual model parameters. Comparing the error estimates from \cref{eq:mc_pval_err,eq:ns_pval_err}, we find a ratio of
\begin{equation}\label{eq:pval_err_ratio}
    \frac{\text{Evaluations for NS}}{\text{Evaluations for MC}} = \frac{(\log^21/p) / \epsilon}{1 / p}\, .
\end{equation}
For small enough \pvalues, NS has the potential to beat MC by an arbitrarily large margin. The efficiency factor, however, could spoil NS performance in realistic applications
and depends the details of the NS implementation and the problem at hand, including the dimensionality. Because of these complications, we will consider the computational performance for two benchmark cases in \cref{sec:examples}.

We summarize and illustrate the idea of ``nested \pvalue computation'' in comparison to MC in \cref{fig:ill}. We want to emphasize that, in the context of this work, \emph{NS is merely a mathematical tool}. \wordlimit{We do not require the evidence estimate or any aspects of Bayesian model comparison, and we do not assume or exploit any connections between the Bayesian evidence and \pvalue. However, there are clearly underappreciated connections between Bayesian and frequentist computation, on which we comment briefly in \cref{sec:conc}.}

\section{Applications and Performance}\label{sec:examples}

To illustrate how to use our algorithm and to assess its performance, we conduct two case studies. First, we consider the \wordlimit{well-studied} example of multi-dimensional Gaussians.
Second\wordlimit{, as a more realistic HEP example,} we apply our algorithm to a simplified version of a resonance search.

\subsection{Gaussian measurements}\label{sec:gausian_measurements}

\begin{figure}
    \centering
    \includegraphics[width=3.375in]{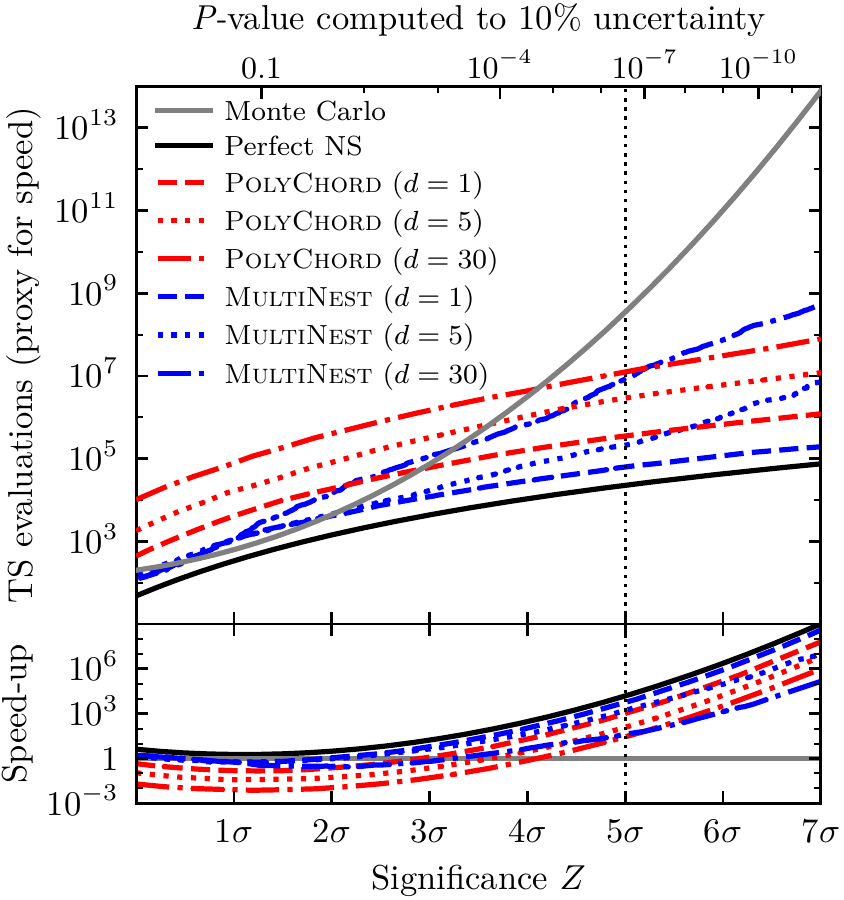}
    \caption{Performance of NS versus MC for the $\chi^2_d$ example. The performance of perfect NS and MC are independent of $d$. Performance was measured by the TS evaluations required to compute the \pvalue to within $10\%$ uncertainty. \wordlimit{NS is orders of magnitude faster for significances greater than about $4\sigma$ and moderate dimension.}}
    \label{fig:performance}
\end{figure}

Consider the \pvalue associated with $d$ independent Gaussian measurements. We define the TS as the sum of the $d$ associated chi-squared variates. Thus $\lambda \sim \chi^2_d$ and the \pvalue may be computed analytically. We construct the problem in NS by mapping from draws from a $d$ dimensional unit hypercube with coordinates $\param_i \sim \mathcal{U}(0, 1)$, to $d$ draws from a chi-squared distribution via $z_i = F_{\chi^2_1}^{-1}(\param_i)$ for $i = 1, \ldots, d$ such that $z_i \sim \chi^2_1$ and $\lambda = \sum z_i \sim \chi^2_d$.

We show the performance of NS and MC on this problem in \cref{fig:performance} as a function of the significance. The performance of perfect NS (in which we suppose that one could sample directly from the constrained prior) and MC are independent of $d$. The performance of real implementations of NS however depends on $d$, as they must utilize numerical schemes for sampling from the $d$-dimensional constrained prior. We demonstrate the performance of \MN~\cite{Feroz:2007kg,Feroz:2013hea} and \PC~\cite{Handley:2015fda}, which utilize ellipsoidal and slice sampling, respectively.\footnote{We changed the codes to stop once the required TS threshold was reached. \wordlimit{The number of live points was dictated by the desired uncertainty. In general, we suggest at least about $\nlive = 100$ or at least $\nlive > d$ as the live points guide the sampling from the constrained prior.} While we chose $\nlive$ to achieve the desired uncertainty, we generally suggest to choose $\nlive \geq \mathrm{max}(100,d)$. We set $\texttt{efr} = 0.3$ and $\texttt{num\_repeats} = 5d$ and \texttt{do\_clustering = False}. See the associated codes for our complete settings~\cite{code}.} We use the number of TS evaluations as a measure of performance, which is sensible when the computational cost is dominated by TS evaluations. While this is usually a good assumption in practice, we note that \MN appears to suffer from a greater computational overhead than \PC in our example, possibly due to the linear algebra operations required for clustering live points and constructing ellipsoids, especially for $d = 30$ over $5\sigma$. For this case, \MN was about 2,000~times slower than \PC~(with sizeable variability in run times), which significantly exceeds the ratio of TS evaluations needed.

\subsection{Resonance search}\label{sec:higgs}

\wordlimit{Our second example is a simplified version of} The original Higgs discovery by the ATLAS Collaboration~\cite{Aad:2012tfa} \wordlimit{in the diphoton channel. This} is a typical example of a resonance search (testing a spectrum for the presence of a signal above a smooth background), which includes the common complications of parameters on the boundary of the parameter space and the look-elsewhere effect.
Here, the null hypothesis~$H_0$ corresponds to the Standard Model~(SM) \wordlimit{background-only} hypothesis, with a known shape and an unknown nuisance parameter~$b$, the total number of background events. The alternative hypothesis $H_1$ is that of the SM background plus a Higgs boson with a Gaussian signal, with a known width but an unknown mass $m_h$ and an unknown positive signal strength $s \ge 0$.\footnote{In the original analysis, the signal is described by a Crystal~Ball function and the background model is more complex~\cite{Aad:2012tfa}. \wordlimit{We ignore these details to provide an accessible description.}}
The null hypothesis lies at the $s = 0$ boundary.
The data $\sampleParams = \{n_i\}$ consists of the Poisson-distributed observed counts~$n_i$ in the 30~bins shown in Figure~4 of~\cite{Aad:2012tfa}. The TS is the log-likelihood ratio
\begin{equation}\label{eq:ts_llr_higgs}
    \lambda(\sampleParams) = 2\log\left(\frac{\max P(\sampleParams \, |\, b, s, m_h) }{\max P(\sampleParams \, |\, b, s = 0)}\right) \, ,
\end{equation}
where we maximize over all parameters for a given (pseudo-)data set~$\sampleParams$.

In summary, the data space~$\sampleParams$, and hence the pseudo-data sampling space, is 30-dimensional while the null and alternative models have only one and three model parameters respectively.

\begin{figure}
    \centering
    \includegraphics[width=3.375in]{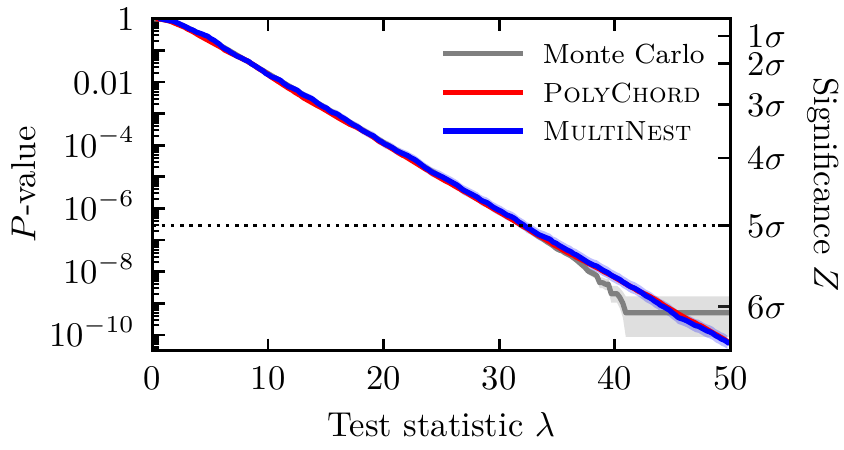}
    \caption{Calibration of the resonance search TS. We show the estimated \pvalue with shaded uncertainties for MC simulations~(gray; \num{2e9} samples), \PC~(red; weighted average of four runs), and \MN~(blue; single run).
    }
    \label{fig:resonance_search}
\end{figure}

To compute \pvalues in the presence of a nuisance parameter~\cite{Demortier:1099967}, we plug in the best-fit value of the unknown nuisance parameter $b$ in \cref{eq:pval_def}. We perform brute-force MC simulations and NS with \MN and \PC to calibrate the TS~$\lambda$. The results are shown in \cref{fig:resonance_search}. \wordlimit{, where we use the conservative Clopper--Pearson intervals~\cite{Clopper1934,10.1214/ss/1009213286} to estimate the MC uncertainty instead of \cref{eq:mc_pval_err}.}

To reach a TS of $\lambda = 50$~($6.5\sigma$) \PC requires about \num{3e6} TS evaluations. \MN typically requires around \num{3.5e7} TS evaluations, i.e.\ an order of magnitude more. \wordlimit{However, this number can sometimes be a factor of a few larger or smaller, due to a less stable performance for \MN, which is not unexpected in higher dimensions.} To achieve a similar level of uncertainty as a single NS run for $\lambda = 50$, where $\Delta\log_{10} p \approx 0.2$ for both \MN and \PC, we would require about \num{e11}~MC~simulations. Since this would be computationally fairly expensive, we only simulated \num{2e9}~samples for \cref{fig:resonance_search}.

The improved performance for NS vs MC in terms of TS evaluations is in line with \cref{fig:performance} and also with CPU run times required: a typical \PC run took about 300\,CPUh~(equivalent hours on a single computing core) while the required \num{e11}~MC~simulations would take around \num{9.2e6}\,CPUh, such that we achieved a speed-up of about 31,000.

Note that the gradient of the linear slope in \cref{fig:performance} also agrees with the asymptotic Gross--Vitells method~\cite{Gross:2010qma}, $\log p \approx c_0 - \frac12 \lambda$. \wordlimit{,
where the unknown constant~$c_0$ must be computed through simulations.} Whilst this method may be used to efficiently compute \pvalues in the resonance search example, it is not applicable in other cases of interest\wordlimit{. When, e.g., the hypotheses under consideration are not nested, one usually has to resort to MC simulations}~\cite{2020NatRP...2..245A}. As a concrete example, consider the goodness-of-fit test of the SM plus Higgs hypothesis, which can be performed using the TS~\cite{Baker:1983tu}
\begin{equation}
    \lambda(\sampleParams) = -2\log\left(\frac{\max P(\sampleParams \, |\, b, s, m_h) }{ P(\sampleParams \, |\, \mu_i = n_i )}\right) \, , \label{eq:higgs_gof}
\end{equation}
with $b$, $s$, and $m_h$ are as in \cref{eq:ts_llr_higgs}. \wordlimit{The difference now is that we normalize the TS by the Asimov data set, i.e.} We set the $\mu_i$ parameters in the Poisson distribution for each bin equal to the pseudo-counts data~$n_i$ in that bin. Since the $\mu_i$ are different from the model parameters of the Higgs hypothesis, these two models are not nested.

The na\"ive expectation for the goodness-of-fit test is that it follows a $\chi_d^2$ distribution with $d = 30 - 3 = 27$ degrees of freedom \wordlimit{~(DOF), subtracting the ``parameter DOF'' from the ``data DOF.''} The implied critical value of the TS for $5\sigma$~($p \approx \num{28.7e-8}$) is then $\lambda^\star \approx 80.8$. However, with direct MC simulations and \PC, we estimate the corresponding \pvalues to be $p = (8.4 \pm 1.0) \times 10^{-8}$ and $p = (6.6 \pm 1.1) \times 10^{-8}$, respectively, or around $5.3\sigma$. This discrepancy from the asymptotic result should be anticipated, as the SM plus Higgs is not a linear model~\cite{andrae2010dos}. \wordlimit{Our goodness-of-fit study also illustrates the benefits of a proper TS calibration as a $5\sigma$ rejection could have already been claimed for $\lambda \gtrsim 76$ as opposed to $\lambda \gtrsim 80.8$.} The six \PC runs that we used for this estimate took on average about \num{1.3e6} function calls and \num{220}\,CPUh. The CPU time per TS function call is consistent with our MC analysis, where we simulated \num{8.2e8} samples in total. \wordlimit{This indicates again that the TS evaluation time dominates over potential computational overhead of \PC, as expected for a TS in realistic applications.} The achieved speed-up of about~\num{100}  is consistent with \cref{fig:performance}. 

\section{Discussion and conclusions}\label{sec:conc}

In this \textit{Letter}, we propose the use of nested sampling~(NS) for computing \pvalues. Our proposal is compatible with readily available, established implementations of the NS~algorithm. \wordlimit{We demonstrated the validity of our algorithm and estimated its speed and uncertainty. The code to reproduce our results is available at \refcite{code}.}
The most notable advantage of using NS is the order-of-magnitudes speed-up \wordlimit{and increase in accuracy} that can be achieved for small \pvalues of scientific importance in comparison with Monte~Carlo~(MC) simulations.
This advantage can be traced to the fact that, despite their conceptual differences, \pvalues and the Bayesian evidence have one thing in common that makes them difficult to compute: \emph{compression}. For the Bayesian evidence, the compression occurs between prior and posterior distributions,
and is typically enormous if the data favor a narrow region of the parameter space.
For \pvalues, the compression is from the whole sampling space to the region where the test statistic~(TS) is greater than its observed value. 
By definition, the compression is enormous for small \pvalues. 

We can understand the applicability of NS to compression problems in both Bayesian and frequentist settings as follows: simulating from the entire sampling distribution (or the entire prior in the Bayesian setting) is inefficient when the interesting region of sampling space is tiny. Conversely, simulating only from the region of interest (or from the posterior in the Bayesian setting) does not allow reliable inference about its relative size.

Heuristically, it is not surprising that successful strategies simulate from a series of distributions that form a path between e.g.\ the prior and posterior
(see \refcite{10.1214/13-STS465,10.1214/ss/1028905934} for further discussion). 
Since the constrained prior can be related to the \pvalue by \cref{eq:p_value_X}, the sequence of constrained priors in NS naturally forms a path between the entire sampling space and the tiny tail area of interest corresponding to the \pvalue, which makes it particularly well-suited for frequentist computation. Path sampling in this manner is a generalization of importance sampling~\cite{10.1214/ss/1028905934}. See \refcite{2013CoPhC.184.2438W} for related work on computing \pvalues through importance sampling and \refcite{Feroz:2013hea} for an importance sampling algorithm that uses NS draws.


We see three potential drawbacks when using NS in this setting. First, the data space in a realistic problem can be very high-dimensional. For example, a search for new particles at the LHC could look for evidence in a histogram containing 100 bins or more. While some implementations of the NS scheme potentially suffer from poor efficiency in higher dimensions, using e.g.\ NS with slice sampling achieves $\epsilon \propto 1 / d$ behavior and has been applied successfully to problems with hundreds of dimensions~\cite{2020arXiv200412211J}. 

Second, while an NS run with $\nlive$ live points can be efficiently parallelized into in principle as many as $\nlive$ runs with a single live point, those individual runs \wordlimit{cannot be further parallelized and} must proceed linearly.\footnote{The uncertainty of the estimate, however, may \wordlimit{always} be reduced by \wordlimit{simultaneously} running the algorithm multiple times \wordlimit{on more processing units}.} By using exponentially large amounts of computational resources, it is therefore possible to make brute-force MC compute \pvalues faster than any realistic NS algorithm, albeit at significantly greater overall computational expense. 

Lastly, there is a subtlety concerning substantial plateaus in the TS, that is, regions of sampling space that lead to an identical TS. In some NS implementations plateaus lead to faulty estimates of the volume variable~\cite{Fowlie:2020gfd}, and thus would lead to faulty estimates of the \pvalue. In such cases, an implementation of NS that supports plateaus must be used or the \pvalue must be corrected using e.g., \textsc{anesthetic}~\cite{Handley:2019mfs}. 


Despite its drawbacks, the in-principle exponential improvement makes our method a valuable tool---in particular when considering small \pvalues. We demonstrated in practice that our algorithm can significantly reduce the computational burden of \pvalue~calculations with the popular \MN and \PC algorithms. This allows for a straightforward adoption of our algorithm, encouraging more rigorous \pvalue calculations and potentially opening up problems that were previously computationally unfeasible.

\wordlimit{On a more conceptual note, our algorithm links Bayesian and frequentist statistics through the common computational challenge of compression, and their common solution of utilizing a bridging sequence of distributions. The NS algorithm is particularly appropriate for this task due to a connection between the constrained prior and the \pvalue.  
While NS is merely a mathematical tool for our purposes, the existence of such a useful connection between Bayesian and frequentist methods hidden in plain sight may inspire further research and potentially uncover more connections between computational methods in the two rivaling statistical interpretations.}

\section*{Acknowledgments}

AF was supported by an NSFC Research Fund for International Young Scientists grant 11950410509. SH was funded by the Alexander von Humboldt Foundation and the German Federal Ministry of Education and Research. WH was supported by a Royal Society University Research Fellowship. We used the Scientific Computing Cluster at GWDG, the joint data centre of Max Planck Society for the Advancement of Science~(MPG) and the University of G\"ottingen. This work was also performed using the Cambridge Service for Data Driven Discovery (CSD3), part of which is operated by the University of Cambridge Research Computing on behalf of the STFC DiRAC HPC Facility (www.dirac.ac.uk). The DiRAC component of CSD3 was funded by BEIS capital funding via STFC capital grants ST/P002307/1 and ST/R002452/1 and STFC operations grant ST/R00689X/1. DiRAC is part of the National e-Infrastructure

\bibliography{references}

\end{document}